\documentclass[aps,prb,twocolumn,superscriptaddress,amsmath,amssymb,showpacs]{revtex4}
\usepackage{amssymb}
\usepackage{graphicx}
\usepackage[dvips]{color}
\usepackage{bm}
\usepackage{epsfig}
\usepackage[latin1]{inputenc}

\begin{document}

\title{La-dilution effects in TbRhIn$_5$ antiferromagnet}
\author{R. Lora-Serrano}
\affiliation{Instituto de F\'isica
"Gleb Wataghin",UNICAMP,13083-970, Campinas-São Paulo, Brazil.}

\author{D. J. Garcia}
\affiliation{Instituto de F\'isica "Gleb Wataghin",UNICAMP,13083-970, Campinas-São Paulo, Brazil.}
\affiliation{Centro Atómico Bariloche, Comisión Nacional de Energía Atómica, 8400 S.C. de Bariloche, Argentina}

\author{E. Miranda}
\affiliation{Instituto de F\'isica "Gleb
Wataghin",UNICAMP,13083-970, Campinas-São Paulo, Brazil.}

\author{C. Adriano}
\affiliation{Instituto de F\'isica "Gleb
Wataghin",UNICAMP,13083-970, Campinas-São Paulo, Brazil.}

\author{C. Giles}
\affiliation{Instituto de F\'isica "Gleb
Wataghin",UNICAMP,13083-970, Campinas-São Paulo, Brazil.}

\author{J. G. S. Duque}
\affiliation{Instituto de F\'isica "Gleb
Wataghin",UNICAMP,13083-970, Campinas-São Paulo, Brazil.}

\author{P. G. Pagliuso}
\affiliation{Instituto de F\'isica "Gleb
Wataghin",UNICAMP,13083-970, Campinas-São Paulo, Brazil.}

\date{\today}

\begin{abstract}
We report measurements of temperature dependent magnetic
susceptibility, resonant x-ray magnetic scattering (XRMS) and heat
capacity on single crystals of Tb$_{1-x}$La$_x$RhIn$_5$ for
nominal concentrations in the range 0 $\leqslant$ \textit{x}
$\leqslant$ 1.0. TbRhIn$_5$ is an antiferromagnetic (AFM) compound
with $T_{N} \approx$ 46 K, which is the highest $T_{N}$ values
along the $R$RhIn$_{5}$ series. We explore the suppression of the
antiferromagnetic (AFM) state as a function of La-doping
considering the effects of La-induced dilution and perturbations
to the tetragonal crystalline electrical field (CEF) on the long
range magnetic interaction between the Tb$^{3+}$ ions.
Additionally, we also discuss the role of disorder. Our
results and analysis are compared to the properties of the undoped
compound and of other members of the $R$RhIn$_{5}$ family and
structurally related compounds ($R_{2}$RhIn$_{8}$ and
$R$In$_{3}$). The XRMS measurements reveal that the commensurate
magnetic structure with the magnetic wave-vector
(0,$\frac{1}{2}$,$\frac{1}{2}$) observed for the undoped compound
is robust against doping perturbations in
Tb$_{0.6}$La$_{0.4}$RhIn$_5$ compound.
\end{abstract}

%\keywords{}

\maketitle


\section{Introduction}

The magnetic dilution and percolation problems are directly
connected to each other and continue to be attractive subjects in
the field of magnetism and strongly correlated electrons systems
(SCES). This is because different and interesting ground states (GS) can be generally tuned by chemical
substitution in the these systems. In particular, for heavy-fermions compounds,
chemical substitution is a very important tuning parameter as it
may strongly affects the interplay between the intra-site Kondo
effect and the inter-site long range
Ruderman-Kittel-Kasuya-Yoshida (RKKY) magnetic interaction,
driving the system from a magnetic ordered state (for instance, antiferromagnetic (AFM) to a non-magnetic heavy-electron
paramagnetic metal.\cite{review} Interestingly, in the vicinity of
the magnetic phase, unconventional superconductivity (USC) and
non-fermi-liquid behavior (NFL) may be found in many
cases. In terms of dilution at the heavy fermions ion
site, a non-obvious evolution from a magnetic or non-magnetic
dense Kondo lattice state to a Kondo single impurity regime in the
very diluted regime is expected.\cite{review}

The family of heavy fermions  Ce$_mM_n$In$_{3m+2n}$ (\textit{M} = Co, Rh or Ir, \textit{m} = 1, 2; \textit{n} = 1) have been proving
to be a great series to explore the role of doping in tuning a variety of ground states such as AFM, USC, NFL and Fermi Liquid (FL) behavior in high-quality single-crystals. All these interesting GS have been found in theses systems in specific
regions of their rich phases
diagrams.\cite{pagliuso1,pagliuso2,Fisk5,Hering,bauer2,Zapf,Moreno}

Dilution studies\cite{pagliuso7,VictorCorrea,Light,Alver,christianson2,wei3,Tanatar,Petrovic3,Satoro} in the above series were performed for both ambient pressure AFM (CeRhIn$_5$) and USC (CeCoIn$_5$) heavy-fermion compounds. In
terms of suppression of AFM, a critical La-concentration of about
$x_c$ = 0.4 was obtained from the extrapolation of $dT_N/dx$ slope
to $T\rightarrow$ 0 for La-doped Ce$_{1-x}$La$_{x}$RhIn$_5$.\cite{pagliuso7} This is
consistent with the theoretical percolation threshold for a
2D-spin system.\cite{Kato} On the other hand,
measurements of thermal expansion and
magnetostriction in Ce$_{0.6}$La$_{0.4}$RhIn$_{5}$ single crystals\cite{VictorCorrea} suggested the
evolution of the crystalline electrical field (CEF) ground state
as a function of La-concentration and revealed the presence of
remaining anisotropic short-range magnetic correlations, which was
consistent with earlier reported heat capacity
data.\cite{pagliuso7,Light} In terms of La-doping induced changes
in the electronic structure, recent de Haas-Van Alphen (dHvA)
measurements in Ce$_{1-x}$La$_{x}$RhIn$_{5}$ has shown a near
insensitivity of the Fermi surface topology to $x$ implying almost
entirely localized $f$-electron behavior.\cite{Alver} The magnetic
structure of the CeRhIn$_{5}$ is also nearly unaffected by 10\% of
La substitution.\cite{wei3}

More recently, studies of Ce$_{0.9}$La$_{0.1}$RhIn$_{5}$ under
pressure have revealed that the La-doping shifts the pressure
induced superconducting phase to higher pressures, indicating that
the main effect of the La-doping in CeRhIn$_{5}$ in this range of
La concentration is the decreasing of the Kondo coupling.\cite{Leticie}

Regarding the effect of La-dilution in the properties of the
superconducting and dense Kondo lattice CeCoIn$_{5}$, the
pair-breaking by non-magnetic La results in a depression of
$T_{c}$ that extrapolates to zero for a critical La-concentration
$x_c$ $\approx 0.18$ indicating a strong gap
anisotropy.\cite{Petrovic3} Further, thermal conductivity and
specific heat experiments at low temperature revealed that the suppression
of $T_{c}$ is followed by the increase in the residual electronic
specific heat but along with the decrease in the residual electronic
thermal conductivity. This contrasting
result suggests a coexistence between unpaired electrons and nodal
quasiparticles.\cite{Tanatar} Still in the Ce$_{1-x}$La$_{x}$CoIn$_{5}$ series, an interesting evolution of the normal state properties was also verified through the finding of scalings laws for the specific heat and magnetic susceptibility data suggesting
two separated energy scales: one from a single-impurity Kondo temperature $T_{K}$ and the other from a larger inter-site spin-liquid
temperature $T^{*}$ which involves the inter-site antiferromagnetic correlations.\cite{Satoro} From their high-$T$ heat capacity data, they claimed that the CEF scheme remains unchanged as a function of La-concentration.\cite{Satoro}

However, to achieve a complete microscopic understanding of the
evolution of physical properties induced by La-doping in
CeRhIn$_{5}$ and CeCoIn$_{5}$ is a very difficult task as the
doping may affect simultaneously, and in a combined way, the
in-site Kondo effect, the inter-site RKKY interaction, the CEF
effects, the electronic structure and also introducing disorder. In this sense, the study of structurally-related
compounds
within the $R_mM_n$In$_{3m+2n}$ family have been successfully used
to understand the evolution of $4f$-electrons magnetism for many
members of the series in situations where some of the
contributions above can be negligible.\cite{pagliuso4,pagliuso3,granado1,granado2,raimundo1,raimundo2,raimundo3,pagliuso5} For instance, in the
Gd$_mM_n$In$_{3m+2n}$ (\textit{M} = Rh and Ir) compounds, as Gd$^{3+}$ is a pure (S = 7/2,
L = 0) spin ion, the RKKY interaction and its dependence with
electronic structure is the main
contribution.\cite{pagliuso3,granado1,granado2} For the Nd- and Tb-based members of the $R_mM_n$In$_{3m+2n}$ family,\cite{pagliuso4,raimundo1,raimundo2,pagliuso5,Chang} both RKKY interaction and CEF effects are present, and the CEF
contribution can be evaluated for Krammers (Nd$^{3+}$, J = 9/2)
and non-Krammer ions (Tb$^{3+}$, J = 6) without the complexity of
the Kondo lattice behavior of the Ce-based compounds.

Nonetheless, to the best of our knowledge, none of these isostructural
magnetic relatives of Ce\textit{M}In$_{5}$ had their properties
systematically investigated as a function of dilution.

In this work, we have studied dilution effects on TbRhIn$_5$ as
the Tb$^{3+}$ ions are substituted by non-magnetic La$^{3+}$ ions
for 0 $\leqslant$ \textit{x} $\leqslant$ 1.0. The TbRhIn$_5$
intermetallic compound\cite{raimundo2} orders
antiferromagnetically with a commensurate magnetic structure
(0,$\frac{1}{2}$,$\frac{1}{2}$) below $T_N$ $\sim$ 46 K, which is
the highest $T_N$ among the $R$RhIn$_5$ compounds. Results from
magnetic susceptibility and specific heat data taken below $\sim$
150 K down to 2 K in Tb$_{1-x}$La$_x$RhIn$_5$ (\textit{x} =
0.15, 0.4 and 0.5) as well as the magnetic structure determination
for the Tb$_{0.6}$La$_{0.4}$RhIn$_5$ compound using resonant x-ray
magnetic scattering (XRMS) are reported. From the analysis of the
evolution of magnetic properties of the studied samples as a
function of La-concentration, we evaluate the role of the
different mechanisms for the suppression of the long-range AFM
coupling by considering dilution, changes in the CEF scheme and
the introduction of disorder. Additionally, the XRMS measurements has shown that the
commensurate magnetic structure (0,$\frac{1}{2}$,$\frac{1}{2}$) observed for the
undoped compound is robust against doping perturbations,
indicating that no changes in the relative spin interaction of
neighboring Tb-spins are taking place.  These results are discussed in a broader
prospective considering others member of the $R_mM_n$In$_{3m+2n}$
family.

\section{Experimental}

All measurements were taken on single-crystalline samples grown by
the Indium excess flux.\cite{Fisk2} Typical crystal sizes were 0.5
cm x 0.5 cm x few mms. The tetragonal HoCoGa$_5$-type structure and
crystals phase purity were confirmed at ambient temperature by X-ray powder diffraction.
Magnetization measurements were performed as a function of temperature in a commercial superconducting quantum interference device (SQUID)
magnetometer. Specific heat data was taken using a commercial physical property measurement system (PPMS) using the
adiabatic relaxation method in the temperature range between 1.9 -
150 K. The x-ray resonant magnetic scattering (XRMS) experiments
were performed at the XRD2 beamline of the Laborat\'orio Nacional
de Luz S\'incrotron (LNLS), Brazil, and the description of the
experimental setup used can be found on Refs. \onlinecite{raimundo2,granado1,granado4}.

\section{Results and Discussion}

\begin{figure}
\centering
        \includegraphics[width=0.4\textwidth]{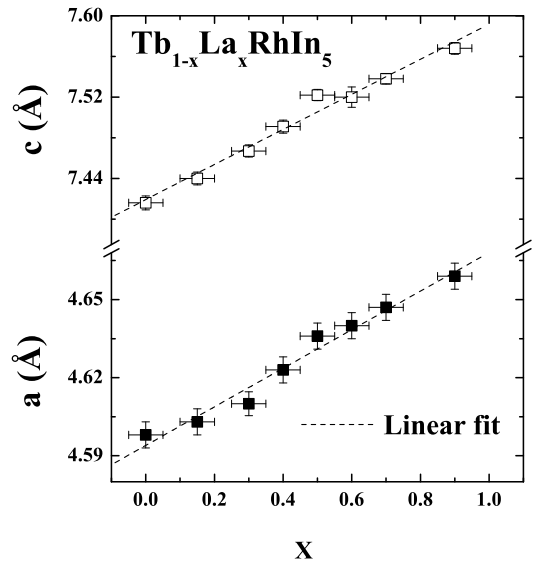}
    \caption{Lattice cell parameters \textit{a} and \textit{c} vs. lanthanum concentration \textit{x} for the Tb$_{1-x}$La$_x$RhIn$_5$ system. The dotted line is a linear fit to both datasets.}
    \label{fig:FigCellParam}
\end{figure}

Fig.~\ref{fig:FigCellParam} displays the cell parameters for
\textit{x} = 0, 0.15, 0.3, 0.4, 0.5, 0.6, 0.7 and 0.9 Lanthanum concentration. Both
parameters expand linearly with concentration \textit{x} (taking
$x$ as the nominal concentration given by the Ce/La ratio in the
starting materials) as the larger La$^{3+}$ ion is substituted
into the Tb$^{3+}$ site, in agreement with the Vegard's law. The
cell parameters \textit{a} and \textit{c} were determined from
least-squares fits of the Bragg peak positions (2$\theta$).\cite{Holland}

\begin{figure}
\centering
        \includegraphics[width=0.48\textwidth]{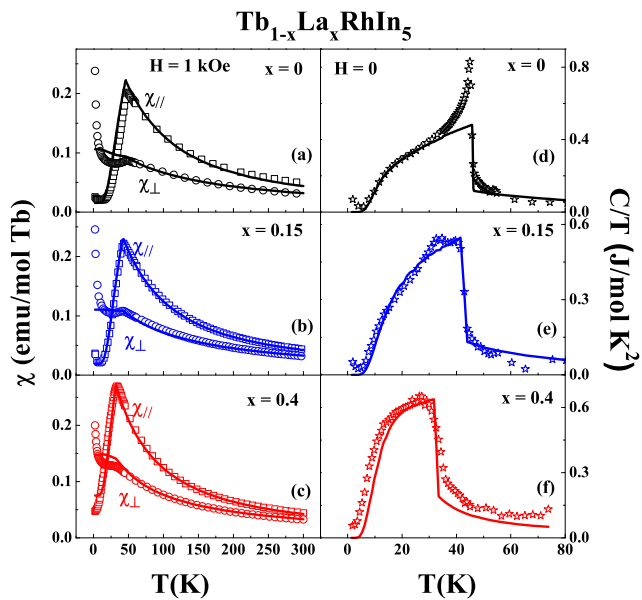}
    \caption{(Color online) (a)--(c) temperature dependence of the magnetic susceptibility for applied field of 1 kOe parallel to \textit{ab}-plane ($\chi_{\bot}$ and circle symbols) and parallel to the [001] direction ($\chi_{//}$ and square symbols) for La-concentrations \textit{x} = 0,\cite{raimundo2} 0.15 and 0.4, respectively. (d)--(f) specific heat, \textit{C/T}, for H = 0 applied field. The solid curves for all cases are the best fits to the data using a MF model.\cite{pagliuso5}}
    \label{fig:FigChiCp}
\end{figure}

Fig.~\ref{fig:FigChiCp} shows the magnetic susceptibility,
$\chi(T)$, and heat capacity, $C(T)/T$, data for representative
samples of Tb$_{1-x}$La$_x$RhIn$_5$. $\chi(T)$ data
[Fig.~\ref{fig:FigChiCp} (a--c)] were taken at a magnetic field
$H$ = 1 kOe applied parallel to the [100] crystallographic
direction ($\chi_{\bot}$) and along the \textit{c} axis ([001]
direction), $\chi_{//}$. Fig.~\ref{fig:FigChiCp} (d--f) display
the temperature dependence of the magnetic specific heat per Tb
mole. The phonon contribution to the total specific heat was
subtracted using the data of non-magnetic YRhIn$_5$. The solid
curves in Fig.~\ref{fig:FigChiCp} are the best fits to the data
using the mean field (MF) model of Ref. \onlinecite{pagliuso5}
which includes an isotropic exchange between rare earth ions and
the tetragonal crystal field terms into the hamiltonian.

The actual La-concentration in our samples was estimated from linear fits to the inverse of the magnetic susceptibility at high-$T$ ($T >$
200 K) assuming the full moment of 9.72 $\mu _B$ for the free
Tb$^{3+}$ ion. The obtained concentrations were found to be in
agreement with the nominal concentration within $\sim$ 4\% for all
doped samples (horizontal error bars in
Fig.~\ref{fig:FigCellParam}). Therefore, we have used the nominal
concentrations, \textit{x}, in this work.

The Fig.~\ref{fig:FigChiCp} demonstrate the shift to lower
values of the temperatures at which the maximum in the
susceptibility occurs, and the specific heat has a peak, as the La-content increases. These
temperatures taken from both measurements coincides reasonably
well, therefore we take this temperature as the N\'eel
temperature, $T_N$, for all samples. From this consideration we
define $T_N$ for the two doped samples $x$ = 0.15 and 0.40 as
being 43 and 34 K, respectively. The shift of $T_N$ to lower
values is a signature from the expected suppression of the
long-range-ordered AFM state. For all cases, the susceptibility is
anisotropic but the ratio $\chi _{//}/\chi _{\bot}$, defined at
the maximum of the $\chi _{//}$ data, remains almost the same
(roughly 2.12, 2.06 and 1.91 for \textit{x} = 0, 0.15 and 0.4,
respectively). Additionally, the transitions in the specific heat
data become evidently broader as a function of La-doping.

\begin{figure}
\centering
        \includegraphics[width=0.4\textwidth]{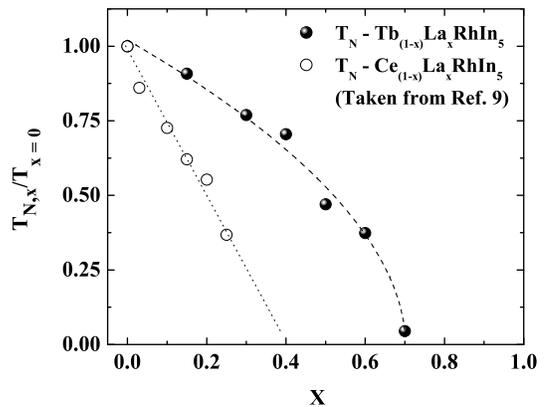}
    \caption{Normalized N\'eel temperature $T_N$ ($T_{N,x}/T_{N(x=0)}$) vs. \textit{x} for Tb$_{1-x}$La$_x$RhIn$_5$ (filled circles)
    determined from the specific heat \textit{C(T)/T} data. The same data for Ce$_{1-x}$La$_x$RhIn$_5$\cite{pagliuso7} (open circles)
    is included for comparison. The dotted line represents linear fit to the Ce-based family data while the dashed curve is
    a power-law fit to the Tb-based data. Extrapolation to $T \rightarrow$ 0 for the Tb$_{1-x}$La$_x$RhIn$_5$ data gives a critical concentration of about 70 \%.}
        \label{fig:TnEvolution}
\end{figure}

Fig.~\ref{fig:TnEvolution} displays the $T_N$ behavior for the
studied compounds normalized by the $T_N$ value of the TbRhIn$_5$
compound ($T_{N,x}/T_{N(x=0)}$) - filled symbols. Similar data
obtained for Ce$_{1-x}$La$_x$RhIn$_5$\cite{pagliuso7} (open
symbols) is included for comparison. Interestingly, the
suppression of $T_N$ as a function of La-doping is less dramatic
in Tb$_{1-x}$La$_x$RhIn$_5$ when compared to
Ce$_{1-x}$La$_x$RhIn$_5$, and its behavior follows approximately
a power-law decrease, differently from the Ce-based series, where
a linear decrease of $T_N$ was observed. The critical
concentration for which $T_N$ $\rightarrow$ 0 was found to be
$x_c$ $\approx$ 0.7 for Tb$_{1-x}$La$_x$RhIn$_5$ in contrast to
the $x_c$ $\approx$ 0.4 found for Ce$_{1-x}$La$_x$RhIn$_5$.

\begin{figure}
\centering
        \includegraphics[width=0.4\textwidth]{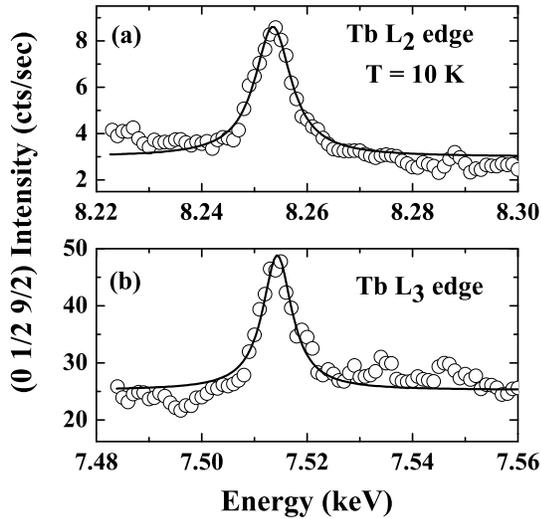}
    \caption{Energy dependence of the resonant x-ray diffraction signal in Tb$_{0.6}$La$_{0.4}$RhIn$_5$ around the (0,$\frac{1}{2}$,$\frac{9}{2}$) satellite peak. (a) Data collected around the Tb $L_2$ absorption edge (8.252 keV) at the base temperature (\textit{T} = 11 K). (b) Scattered signal around the Tb $L_3$ edge (7.514 keV). From a single Lorentzian-profile fit to both datasets (continuous line) we extracted the width of the resonance through the full width at half maximum.}
            \label{fig:XRMSEnergyScans}
\end{figure}

La-doping perturbations in the AFM state of TbRhIn$_5$ was further explored through x-ray magnetic diffraction experiments in a crystal of Tb$_{0.6}$La$_{0.4}$RhIn$_5$ from the same batch used for the macroscopic measurements above. These measurements were performed with the incident photon energy at both $L_2$ and $L_3$ Tb absorption edges (resonant condition) in order to enhance the small signal from the AFM order of Tb ions below
$T_N$.\cite{Blume1} We found satellite peaks at reciprocal space positions corresponding to the same reciprocal propagation vector found in the undoped TbRhIn$_5$, i.e. (0,$\frac{1}{2}$,$\frac{1}{2}$),\cite{raimundo2} indicating that Tb$_{0.6}$La$_{0.4}$RhIn$_5$ orders in a commensurate AFM single
\textbf{k} structure (\textbf{k} - propagation vector). Above $T_N$ we only found charge Bragg peaks from the tetragonal HoCoGa$_5$-type structure. Other magnetic peaks, representing twinned AFM domains, were also observed at ($\frac{1}{2}$,0,$\frac{9}{2}$), ($\frac{1}{2}$,0,$\frac{11}{2}$)
and ($\frac{1}{2}$,0,$\frac{13}{2}$) reciprocal space positions (not shown). A comparison between the intensities of the symmetrically-equivalent reflections ($\frac{1}{2}$,0,$\frac{9}{2}$) and (0,$\frac{1}{2}$,$\frac{9}{2}$) reveals a higher (0,$\frac{k}{2}$,$\frac{l}{2}$) domain population over the ($\frac{h}{2}$,0,$\frac{l}{2}$) ones [\textit{h, k, l} integers], the later representing roughly 75\% of the former. 

Open circles in Fig.~\ref{fig:XRMSEnergyScans}(a) and (b) represent the resonance profiles of the superlattice diffraction peak (0,$\frac{1}{2}$,$\frac{9}{2}$) around the $L_2$ (8.253 keV) and $L_3$ (7.514 keV) Tb absorption edges, respectively, taken at 11 K. The spectral shapes are typical of magnetic scattering from the ordered moments of the Tb ion sublattices and the peak maxima coincides with the inflection point of the fluorescence spectrum (not shown), revealing the \textit{E}1 electric dipole-type resonance involving electronic transitions 2$p_{1/2}\leftrightarrow$5$d$ and 2$p_{3/2}\leftrightarrow$5$d$. Therefore, we used the energy where maxima in Figs.~\ref{fig:XRMSEnergyScans}
take place as incident energies for all our measurements of magnetic peaks. Full lines are single-Lorentzian fits from which we were able to get the resonance width, as being 8.4 eV for $L_2$ and 6.7 eV for $L_3$. This width is inversely proportional to the resonance core-hole lifetime. The photon energy variation profile at a fixed reciprocal point has proved to be higher ($I(L_{III})/I(L_{II}) \approx$ 3 eV) and narrower at the $L_3$ Tb edge, which confirms the $L$-edge resonances behavior of the Tb$^{3+}$ ion previously suggested in Ref. \onlinecite{Veenendaal} and observed for other rare-earth-based compounds.\cite{raimundo2,Kim}

\begin{figure}
\centering
        \includegraphics[width=0.45\textwidth]{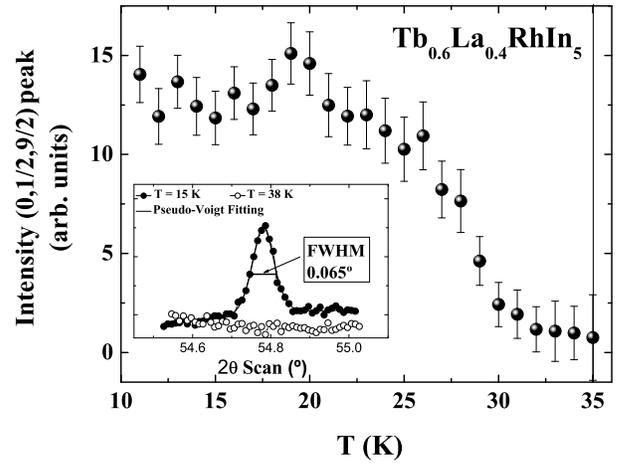}
    \caption{Temperature dependence of the integrated intensities of the (0,$\frac{1}{2}$,$\frac{9}{2}$) magnetic reflection in the
temperature range between \textit{T} = 11 K and 37 K for Tb$_{0.6}$La$_{0.4}$RhIn$_5$. The inset shows two longitudinal scans ($\theta - 2\theta$) around the (0,$\frac{1}{2}$,$\frac{9}{2}$) position: open circles represent the data obtained at \textit{T} = 15 K and the filled circles curve was taken at \textit{T} = 38 K. Continuous line is a Pseudo-Voigt fit to the observed data at \textit{T} = 15 K and the vertical dashed line at \textit{T} = 35 K shows the N\'eel temperature, above which no further long-range order can be found.}
            \label{fig:FigMagScatTDep}
\end{figure}

Fig.~\ref{fig:FigMagScatTDep} displays the temperature dependence
of the (0,$\frac{1}{2}$,$\frac{9}{2}$) magnetic Bragg peak
intensities, which is proportional to the Tb magnetization
sublattice, obtained from numerical integrations to $\theta -
2\theta$ scans (using a Pseudo-Voigt function). The data was taken
between T = 11--37 K with a \textit{T}-step of 1 K while warming
the sample. The inset shows the experimental (filled circles at
\textit{T} = 15 K and open ones at \textit{T} = 37 K) and
calculated curves (continuous line) together. Error bars in the
main panel represent statistical standard deviation from the fits.
The decrease of the Bragg intensities as the temperature is
increased toward the bulk $T_N$ denotes also the magnetic
character of this reflection. It smoothness is a signature of a
second order-type transition.

The XRMS results, together with the properties shown in
Fig.~\ref{fig:FigChiCp}, demonstrate the existence of long-range
AFM correlations for the \textit{x} = 0.4 doped sample. From the
point of view of magnetic diffraction, it seems that the AFM
propagation vector does not change as function of La-doping up to
$x$ = 0.4. As such, we may argue that the relative spin
orientation of neighboring Tb$^{3+}$ ions is not strongly modified
by dilution in the x = 0.4 sample, re-enforcing the long range
character of the RKKY interaction between the Tb$^{3+}$ ions.
Further, this result may be indicative that the balance between
the Tb first and second-neighbors interactions ($J_1$ and $J_2$, respectively
)
is the same as for the undoped TbRhIn$_5$
compound.\cite{granado2,raimundo2}

\begin{table*}[t]
\caption{$T_N$ and CEF parameters for Tb$_{1-x}$La$_x$RhIn$_5$} \label{tab:table1}
\begin{ruledtabular}
\begin{tabular}{ccccccccc}
 & $T_N$(K) & J$_{R-R}$(K) & $B_{20}$(K) & $B_{40}$(K) & $B_{44}$(K) & $B_{60}$(K) & $B_{64}$(K)
\\ \hline TbRhIn$_5$ & 45.6 & 1.9 & -2.2 & 6.6 x 10$^{-3}$ & 6.4 x 10$^{-2}$ & -4.5 x 10$^{-5}$ & 2.7 x 10$^{-3}$

\\ \hline Tb$_{0.85}$La$_{0.15}$RhIn$_5$ & 42.8 & 1.8 & -1.8 & 6.8 x 10$^{-3}$ & 5.6 x 10$^{-2}$ & -7.5 x 10$^{-5}$ & 2.3 x 10$^{-3}$

\\ \hline Tb$_{0.6}$La$_{0.4}$RhIn$_5$ & 32 & 1.6 & -1.8 & 1.1 x 10$^{-2}$ & -3.6 x 10$^{-3}$ & -1.0 x 10$^{-4}$ & -2.6 x 10$^{-3}$
\end{tabular}
\end{ruledtabular}
\end{table*}

Nonetheless, from our recent data we can not determine the
direction of magnetic moments in the Tb sub-lattice through the
comparison between observed and calculated integrated intensities
of magnetic peaks because only three reflections from the same AFM
domains were reached with our experimental setup. Therefore, new
data in the resonant condition are required to know the moments
orientation for this La-doped sample. Particularly, it should be
interesting to includes azymuthal dependence of magnetic peaks
intensity combined with polarization analysis.

\begin{figure}
\centering
        \includegraphics[width=0.45\textwidth]{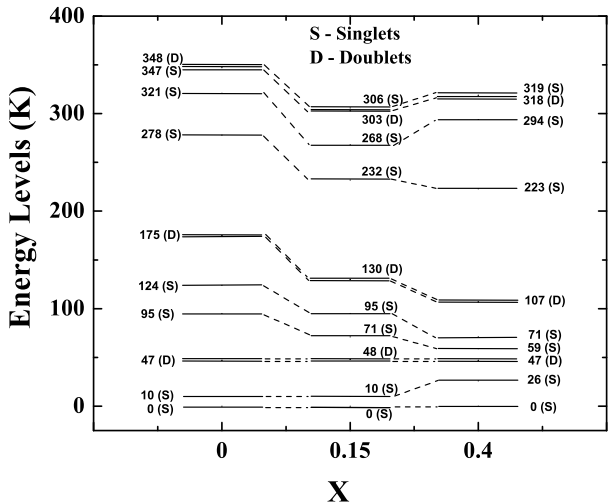}
    \caption{CEF splitting of the ground state multiplet of Tb obtained from
 the simulations of Fig.~\ref{fig:FigChiCp} to the \textit{x} = 0,
0.15 and 0.4 using the MF model of Ref.
\onlinecite{pagliuso5}}\label{fig:CEFScheme}
\end{figure}

We now discuss the effects of La-doping in the antiferromagnetic
interaction between the Tb$^{3+}$ ions in TbRhIn$_{5}$. The first
obvious effect is dilution. As La$^{3+}$ replaces the Tb$^{3+}$,
the average distance between the remaining Tb$^{3+}$ ions increases
and consequently the RKKY magnetic exchange between them decreases. Secondly, there is the chemical pressure effect
induced by the difference in ionic size between La$^{3+}$ and
Tb$^{3+}$. This can affects the CEF effects at the Tb$^{3+}$ site.
These effects maybe subtle but are not straightforward. Slight
modifications in the CEF scheme and/or wave-functions can cause
significant changes in $T_N$ and in the magnetic anisotropy of the
ordered state for low-symmetry systems.\cite{pagliuso5} Last, there could exist the effect of chemical disorder caused by a not
perfectly homogeneous La-distribution through the sample. This may
cause competing magnetic interaction between Tb$^{3+}$ ions in
different grains, creating multiple spin configuration and/or
distribution of $T_N$, leading to the suppression of the long range ordered state.

In order to account for the evolution of the first and second
effects above we have used our MF model from Ref.
\onlinecite{pagliuso5} to fit concomitantly the whole set of data
of Fig.~\ref{fig:FigChiCp}. In Fig.~\ref{fig:CEFScheme} we show
the CEF schemes obtained from the best fits to the data of the representative samples shown in Fig.~\ref{fig:FigChiCp}.

Before we proceed with the analysis of the results presented in
Fig.~\ref{fig:CEFScheme} it is important to discuss the
reliability of these results because CEF parameters obtained from
fits to macroscopic measurements data could be mistaken and
not unique. Is is known that, in general, a given set of
parameters can describe very nicely a set of experimental
macroscopic results and completely fail in describe others and that a
definitive determination of CEF schemes and/or parameters usually
requires direct measurements by inelastic neutron scattering (INS).

\begin{figure}
\centering
        \includegraphics[width=0.47\textwidth]{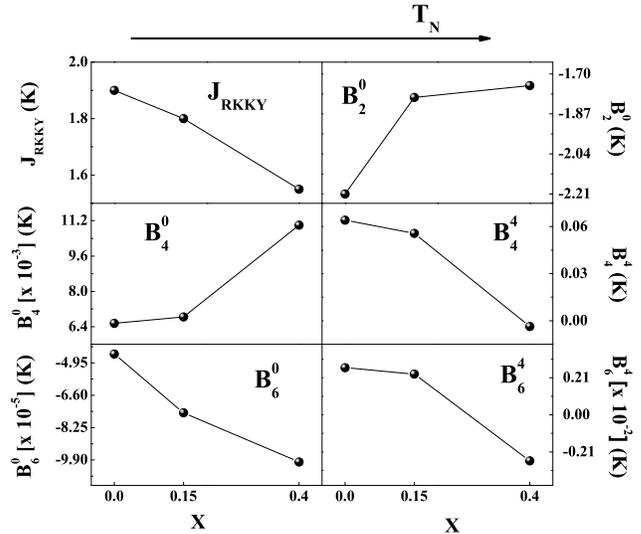}
    \caption{CEF parameters evolution as the La-concentration is increased.}
            \label{fig:CEFParam}
\end{figure}

In an earlier report, we proposed a CEF scheme for pure
TbRhIn$_{5}$ obtained from fits to magnetic susceptibility and
specific heat data. This CEF scheme was based on a
$\Gamma_{5}^{(1)}$ doublet ground state and an overall splitting
of 310 K.\cite{raimundo2} Later, low temperature magnetization
experiments\cite{Takeuchi2007} shown that this scheme is
incompatible with the high field ($H$ $>$ 10 T) magnetization data
taken at $T$ = 2 K for a magnetic field applied along the $c$-axis.
As such, in Ref. \onlinecite{Takeuchi2007}, the authors proposed
an alternative scheme of levels with a singlet ground state and overall splitting of about 220 K. This alternative scheme agrees
qualitatively with the high field behavior of their $T$ = 2 K magnetization data but it does not give a better fit to the $\chi(T)$ and zero field $C_{p}(T)$ data than that obtained with the CEF scheme of
Ref. \onlinecite{raimundo2}.

Taking into account all the above, we have re-analyzed our
$\chi(T)$ and zero field $C_{p}(T)$ and the high field
magnetization data of Ref. \onlinecite{Takeuchi2007} and obtained
 the new CEF scheme of levels present in Fig.~\ref{fig:CEFScheme} for pure
 TbRhIn$_{5}$. The new scheme has a singlet ground-state with a first excited singlet at 10 K and an overall splitting of $\sim$ 350 K (see
Fig.~\ref{fig:CEFScheme}, left scheme). Although both present a singlet ground state, the TbRhIn$_{5}$ CEF scheme of Fig.~\ref{fig:CEFScheme} and the one from Ref. \onlinecite{Takeuchi2007} display appreciable differences in terms of energy level and wave functions. These level of controversy usually requires direct measurements by INS to be completely solved, however our preliminary CEF parameters determination and data fits proceeding allow us to follow the La-doping induced changes in the crystal field and the modifications on the AFM state of TbRhIn$_{5}$. Further, it is
important to emphasized that the qualitative evolution of the CEF scheme and parameters as a function of La-doping was found to be nearly
independent of the details of a particular choice of CEF parameters (and scheme) for pure TbRhIn$_{5}$ (Fig.~\ref{fig:CEFScheme} or Ref. \onlinecite{Takeuchi2007}).

Analyzing the La-doping evolution of the CEF energy level schemes
in Fig.~\ref{fig:CEFScheme}, we observe that for $x$ $\lesssim$
0.4 the best fits yield ground-state singlets and a non-monotonic evolution of the overall
CEF splitting as a function of La-doping. However, one can clearly
observes a trend in the low-$T$ energy levels ($T <$ 200 K), showing a compression to lower temperature ranges. This effect causes the increase of the low-$T$ magnetic entropy and may
certainly affect $T_N$ and the ordered state.\cite{pagliuso5}
Additionally, the set of CEF parameters giving place to the fits of Fig.~\ref{fig:FigChiCp}, and to the schemes of
Fig.~\ref{fig:CEFScheme}, show systematic changes as the La-concentration is increased, see Table~\ref{tab:table1} and Fig.~\ref{fig:CEFParam}.

From Fig.~\ref{fig:CEFParam} one can obviously notice the effect of dilution by decreasing the effective Tb$^{3+}$--Tb$^{3+}$ exchange interaction, $J_{RKKY}$, as well as the modification of the crystal field parameters due the Lattice expansion caused by La-doping. Although these two effects are maybe expected, it is not obvious how they are combined to suppress $T_N$ and to affect the details of the ordered state of complex magnetic systems in dilution studies.

\begin{figure}
\centering
        \includegraphics[width=0.45\textwidth]{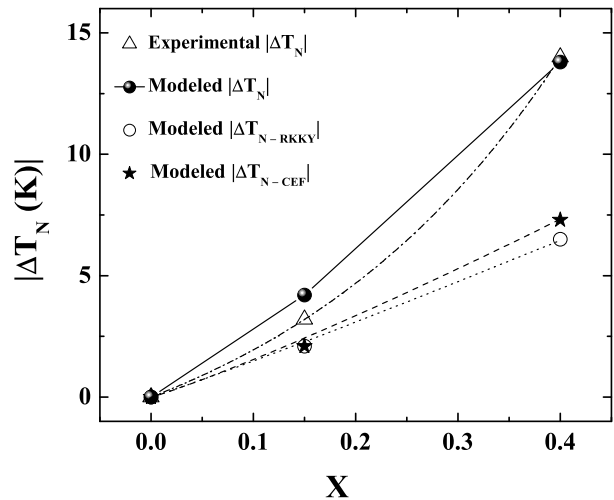}
    \caption{N\'eel Temperature variation, $|\Delta T_N|$, as a function of La-content while changing CEF parameters and $J_{RKKY}$ is fixed to 0.2 meV ($|\Delta T_{N-CEF}|$, star symbols) and as $J_{RKKY}$ was changed with CEF parameters fixed ($|\Delta T_{N-RKKY}|$, open circles symbols). Modeled $|\Delta T_N|$ is the result of considering both effects together.}
            \label{fig:Fig8}
\end{figure}

In order to explore separately the two effects above, we used our MF model to simulate the evolution of $T_N$ while changing the CEF parameters values of Table~\ref{tab:table1} for a fixed $J_{RKKY}$ = 0.2 meV (obtained for the undoped TbRhIn$_{5}$). Alternatively, we study the $T_N$ suppression with the La-doping solely due to the decrease of $J_{RKKY}$ and keeping unchanged the CEF scheme of the TbRhIn$_{5}$ compound. The results of these procedures can be seen in Fig.~\ref{fig:Fig8}. From this plot it is evident that the CEF changes and the decrease of $J_{RKKY}$ have comparable effects on the $T_{N}$ suppression.
The $T_{N}$ shift due the changes of the CEF
parameters by La-doping, $|\Delta T_{N-CEF}|$, is nearly the same as the $T_{N}$ shift resulting from the increase of the
Tb$^{3+}$--Tb$^{3+}$ average distance changes by dilution,
$|\Delta T_{N-RKKY}|$. However,
both effects had to be taken into account to best reproduce the
data of Fig.~\ref{fig:FigChiCp} and the experimental $T_{N}$
suppression of Fig.~\ref{fig:Fig8} (what we called modeled $|\Delta T_N|$). The significant decrease of
$T_{N}$ induced by CEF is a non-trivial and important result that
indicates that the effects of CEF changes should be generally taken
into account when important temperatures scales are mapped and
analyzed as function of doping in the \textit{R}RhIn$_{5}$ family and
related
compounds.\cite{pagliuso1,pagliuso2,Fisk5,Hering,bauer2,Zapf,Moreno,pagliuso7,VictorCorrea,Light,Alver,christianson2,wei3,Tanatar,Petrovic3,Satoro}

Another important aspect of our results for
Tb$_{1-x}$La$_x$RhIn$_5$ is the role of disorder on the
properties of this series. Our theoretical model does not include
any kind of disorder (for instance, a random distribution of
$T_{N}$ and/or CEF parameters) and this is probably the reason why
the fittings curves cannot reproduce the width of the AFM transition in the \textit{C/T} data of Fig.~\ref{fig:FigChiCp} for samples with
higher La-concentration. This also prevents our models to achieve better fits to our data for $x$ = 0.4. On the other hand, it
interesting to notice that we do not see a clear contribution of
the disorder in the suppression of the $T_{N}$ for $x$ $\lesssim$
0.4, as we were able to simulate the experimental data and the $T_N$ behavior only by
considering dilution and CEF tuning effects (see
Figs.~\ref{fig:FigChiCp} and~\ref{fig:Fig8}).

Finally, the last interesting point to be addressed is the
comparison between the $T_{N}$ suppression in
Tb$_{1-x}$La$_x$RhIn$_5$ and in its HF counterpart
Ce$_{1-x}$La$_x$RhIn$_5$ given in Fig.~\ref{fig:TnEvolution}. It
is evident that the critical concentration $x_c$ $\approx$ 0.7 for
Tb$_{1-x}$La$_x$RhIn$_5$ is much higher than the $x_c$ $\approx$
0.4 found for Ce$_{1-x}$La$_x$RhIn$_5$.\cite{pagliuso7} While the $x_c$ $\approx$ 0.4 for Ce$_{1-x}$La$_x$RhIn$_5$ is close to the 2D percolation threshold for a Heisenberg square
lattice,\cite{pagliuso7,Kato} the $x_c$ $\approx$ 0.7 for
Tb$_{1-x}$La$_x$RhIn$_5$ is the same as for a 3D lattice. Thus, a simple argument to
understand the increase in $x_c$ for Tb$_{1-x}$La$_x$RhIn$_5$
may be given by its more 3D character, as the tetragonal lattice
parameters ratio $c/a$ decreases along the \textit{R}RhIn$_5$
series.\cite{pagliuso3,raimundo2} On the other hand, it is clear
that the non-linear suppression of $T_{N}$ is in obvious contrast
to the linear behavior found for Ce$_{1-x}$La$_x$RhIn$_5$ and
even for cubic Ce$_{1-x}$La$_x$In$_3$ ($x_c$ $\approx$ 0.7). In this regard, it is interesting to notice that
for Tb$_{1-x}$La$_x$RhIn$_5$ we found a roughly linear decrease
of $T_{N}$ as a function of $x$ up to $x$ $\sim$ 0.4 (see
Fig.~\ref{fig:TnEvolution}). This linear behavior of the $T_{N}$ decrease could be successfully understood by monotonic evolution
of $J_{RKKY}$ and the CEF parameters as a function of $x$ (see
Figs.~\ref{fig:CEFParam} and~\ref{fig:Fig8}). However, for  0.4 $<$
$x$ $\lesssim$ 0.7, the $T_{N}$ decrease becomes more abrupt and
we were no longer able to fit the data using our model in this
range of concentration, presumably due to the role of disorder.
Therefore, we speculate that the non-trivial $T_{N}$ suppression
for 0.4 $< x \lesssim$ 0.7 in Tb$_{1-x}$La$_x$RhIn$_5$ may
be related to the details of the disorder effects near the
percolation threshold. This effect have not been observed in
Ce$_{1-x}$La$_x$RhIn$_5$ and cubic Ce$_{1-x}$La$_x$In$_3$
because the difference in ionic size between Tb$^{3+}$ and
La$^{(3+)}$ ions is much larger than between Ce$^{3+}$ and
La$^{(3+)}$. In addition, the percolation problem including an
exchange interaction with long range character, as the RKKY
interaction, have not been completely understood even in a
perfectly ordered system.\cite{Hoyos} In fact, our XRMS results for $x$=0.4 indicate that the relative exchange interaction
between neighboring Tb$^{3+}$ ions is very robust against
La-doping which suggests that, due to the long range character of
the RKKY interaction, the dilution-induced $J_{RKKY}$ decrease may be much smaller than that expected for a Heisenberg square
lattice,\cite{pagliuso7,Kato} specially for a more 3D-system.

\section{Conclusions}

In summary, we have performed magnetic susceptibility and specific
heat measurements for the Tb$_{1-x}$La$_x$RhIn$_5$ (\textit{x} =
0.15, 0.3, 0.4 and 0.5) diluted compounds. We also presented preliminary results of resonant x-ray magnetic scattering experiments in Tb$_{0.6}$La$_{0.4}$RhIn$_5$. The AFM structure revealed is commensurate with the same propagation vector (0,$\frac{1}{2}$,$\frac{1}{2}$) of the undoped
compound, which indicates the same relative interaction $J_1$/$J_2$ between Tb$^{3+}$-neighbors. N\'eel temperature decreases with a non-linear behavior as a function of Lanthanum concentration and
extrapolates to zero at roughly 70\% of La content, which demonstrates that for TbRhIn$_5$ the non-magnetic La-substitution shifts the dilution limit
differently to the $x_c$ $\sim$40 \% observed for Ce$_{1-x}$La$_x$RhIn$_5$ and (Ce$_{1-x}$La$_x$)$_2$RhIn$_8$ families. Furthermore, our mean field model simulation for Tb$_{1-x}$La$_x$RhIn$_5$ ($x \lesssim$ 0.4)
reveals that the crystal field scheme evolves as a function of
doping and that this evolution affects $T_{N}$ as much as the
decreasing in $J_{RKKY}$ due to dilution. This effect may be of
great importance in phase diagrams of complex magnetic systems
where the AFM is tuned to zero temperature by chemical
substitution.

\begin{acknowledgments}
This work was supported by FAPESP (SP-Brazil) Grants No. 06/50511-8, 06/60440-0 and 06/60387-2, CNPq (Brazil) Grant No.
305161/2006-7 and Capes (Brazil) Grant No. 065/2007. LNLS is also acknowledged for beamtime at XRD2 beamline.

\end{acknowledgments}

\end{document}